\documentclass[10pt,conference,a4paper]{IEEEtran}
\IEEEoverridecommandlockouts
\usepackage{cite}
\usepackage{amsmath,amssymb,amsfonts}
\usepackage{algorithmic}
\usepackage{graphicx}
\usepackage{textcomp}
\usepackage{xcolor}
\usepackage{subcaption}
\usepackage{microtype}
\usepackage{makecell}
\usepackage{float}
\def\BibTeX{{\rm B\kern-.05em{\sc i\kern-.025em b}\kern-.08em
    T\kern-.1667em\lower.7ex\hbox{E}\kern-.125emX}}

\usepackage{setspace}
\begin{document}

\title{Privacy preserving layer partitioning for Deep Neural Network models\\
}

\author{
\IEEEauthorblockN{Kishore Rajasekar, Randolph Loh, Kar Wai Fok, and Vrizlynn L. L. Thing}
\IEEEauthorblockA{\textit{ ST Engineering} \\
Singapore
}
}

\maketitle

\begin{abstract}
MLaaS (Machine Learning as a Service) has become popular in the cloud computing domain, allowing users to leverage cloud resources for running private inference of ML models on their data. However, ensuring user input privacy and secure inference execution is essential. One of the approaches to protect data privacy and integrity is to use Trusted Execution Environments (TEEs) by enabling execution of programs in secure hardware enclave. Using TEEs can introduce significant performance overhead due to the additional layers of encryption, decryption, security and integrity checks. This can lead to slower inference times compared to running on unprotected hardware. In our work, we enhance the runtime performance of ML models by introducing layer partitioning technique and offloading computations to GPU. The technique comprises two distinct partitions: one executed within the TEE, and the other carried out using a GPU accelerator. Layer partitioning exposes intermediate feature maps in the clear which can lead to reconstruction attacks to recover the input. We conduct experiments to demonstrate the effectiveness of our approach in protecting against input reconstruction attacks developed using trained conditional Generative Adversarial Network(c-GAN). The evaluation is performed on widely used models such as VGG-16, ResNet-50, and EfficientNetB0, using two datasets: ImageNet for Image classification and TON\_IoT dataset for cybersecurity attack detection.

\end{abstract}


\section{Introduction}
MLaaS (Machine Learning as a Service) has become a popular approach to deploy trained deep learning models, provided by cloud service giants like Microsoft Azure, Amazon AWS, and Google Cloud. Users typically send their data such as images and text, to cloud-based MLaaS platforms for inference tasks. 
Trusted Execution Environments (TEEs) such as  Intel® Software Guard Extensions (Intel SGX)\cite{mckeen2013innovative} can be used to preserve the confidentiality of user data. Running full inference of the complete trained model in Intel SGX on encrypted user input which makes it invisible for the cloud service provider. However, the performance gap in running inference on accelerators like GPU compared to running within TEE is very high. Hence, we adopt layer partitioning technique where execution of inference of the trained model is split into critical and non-critical partitions. 
Critical model partition is executed within enclave SGX and non-critical part are the layers offloaded to GPU. While leveraging the GPU can increase computational efficiency, exposing intermediate feature maps in the cloud poses a risk of reconstruction attacks. Our goal is to identify the optimal layer for partitioning and enhance privacy protection.
The following are the contributions made by this paper:
\begin{enumerate}
\item Analyze the inference runtime performance of three image classification models, namely VGG-16, 
ResNet-50, and EfficientNetB0 using layer partitioning techniques for Python workloads within the context of TEE.
\item 
We then measure the efficacy of layer partitioning using trained conditional Generative Adversarial Network (c-GAN) models to evaluate the privacy vs efficiency in the context of the three models. We evaluated ResNet-50 and EfficientNetB0 on two datasets: the ImageNet Kaggle ILSVRC 2012-2017 test dataset \cite{imagenet} and the cybersecurity TON\_IoT dataset \cite{alsaedi2020ton_iot}. 

\item Additionally, we examine whether the choice of dataset influences the reconstruct-ability of input images and also determine if the speedups vary for different models after identifying optimal partitioning points.
\end{enumerate}

\section{Related Works}
\label{sec:related}

Different methods can be employed to safeguard data privacy, including Homomorphic Encryption libraries \cite{gilad2016cryptonets,mishra2020delphi}, Secure Multi-Party Computing\cite{wagh2019securenn} , Differential Privacy\cite{team2017learning}, and the utilization of TEEs. Each approach offers distinct levels of privacy protection and incurs varying costs\cite{mireshghallah2020privacy}. 
 
In the case of TEE-based approaches, the TEE-shielding approach runs the complete unmodified model inside enclaves, ensuring both model confidentiality and high accuracy comparable to the original model. 
The partition-based approach involves manually selecting sensitive model layers to execute within an enclave, while allocating the remaining layers to an untrusted GPU for acceleration. This strategy results in reduced inference latency compared to TEE-shielding approaches. Some prior works include eNNclave\cite{schlogl2020ennclave} and AegisDNN \cite{xiang2021aegisdnn}.
eNNclave replaces partitioned operators' parameters with pre-trained parameters from other publicly available models, which can lead to a loss in inference accuracy. AegisDNN uses dynamic programming to identify partitioning point to learn each layer’s criticality, and partitions uncritical (plaintext) layers to GPU to meet the user argument deadline. 


Slalom\cite{tramr2019slalom} provides an inference framework that uses TEE-GPU collaboration to protect data privacy and integrity. It offloads computational intensive convolutions to GPU, and preserves data privacy of  the offloaded computations using cryptographic blinding technique. 
Origami inference\cite{narra2021origami} uses a combination of model partitioning, computational offloading and eliminates data blinding in second tier of inference to provide fast and private inference by protecting input privacy against a trained c-GAN adversary\cite{mirza2014conditional}.

We aim to evaluate our layer partitioning approach on three models.
To protect exposed intermediate feature maps from input reconstruction attacks, we adopt the method outlined in \cite{narra2021origami}. This involves performing reconstruction attacks using c-GAN models at each partition point. 
We use the trained model with its original parameters and execute non-critical layers of the model in plaintext on GPU. 
Existing approaches depend on the Intel SGX SDK and lack direct support for popular Python deep learning frameworks like PyTorch or TensorFlow. Porting Python models to SGX enclave is difficult.
Gramine \cite{tsai2017graphene},\cite{gram}, on the other hand, enables the execution of unmodified applications within SGX enclaves, eliminating the need for manual porting. 

\section{Research Problem}

\label{sec:research}

\subsection{Intel SGX}\label{AA}
Intel SGX a technology developed by Intel that provides a secure hardware enclave for protecting sensitive data and computations. 
By running trained models within an SGX enclave, user data can be encrypted, decrypted, and processed securely, protecting it from potential threats in untrusted cloud environments. Use of Intel SGX can enhance the privacy and security of machine learning services, enabling users to safely utilize cloud-based inference while preserving the confidentiality of their data.
\subsection{Model Partitioning}
Layer partitioning is a technique that allows for the efficient execution of machine learning model inference by dividing the computational workload between an SGX enclave and a GPU. When running the entire model inference within an SGX enclave, limitations such as memory constraints and performance overhead can arise, especially for large models. To address these challenges, layer partitioning divides the model into two partitions. The first partition consists of the initial layers of the model, which are executed within the SGX enclave. These layers typically contain the majority of the sensitive information that could potentially lead to input reconstruction attacks. User input privacy evaluation is done based on the insight that only the first few layers of the model contain most of the information required for input reconstruction compared to the output from deeper layers of the model.
Within the SGX enclave, the input data is decrypted and processed. The intermediate feature maps resulting from this computation are then offloaded to an untrusted GPU for further processing. The second partition, comprising the remaining layers of the model, can be executed on the GPU for faster execution. This approach optimizes the utilization of resources by leveraging the security of SGX for the critical layers while taking advantage of the computational power of the GPU for the non-critical layers. 
\subsection{Threat model}
We define the adversary as an agent that tries to use observed intermediate feature maps of model to reconstruct input in the untrusted cloud environment. The term `untrusted cloud' refers to cloud environments where the presence of third-party adversaries poses a risk of unauthorized access or observation, potentially enabling them to intercept or monitor sensitive data. 
We use trained c-GAN adversary models for the reconstruction attack evaluation, as outlined in Origami Inference \cite{narra2021origami}, to assess the reconstruct-ability of input images.
The architecture of the c-GAN model contains a Generator and Discriminator. 
The generator follows an encoder-decoder architecture with two residual blocks. 
The discriminator architecture consists of two parts: a down-sampler and a series of convolutional blocks. 
During training, the adversarial BCELoss loss is used to measure the difference between predicted and target labels. 
The training is done for 200 epochs. The reconstructed images are of dimension $3\times224\times244$.
\section{Proposed Solution Framework}
\label{sec:framework}
\begin{figure}[ht]
\centerline{\includegraphics[width=0.47\textwidth]{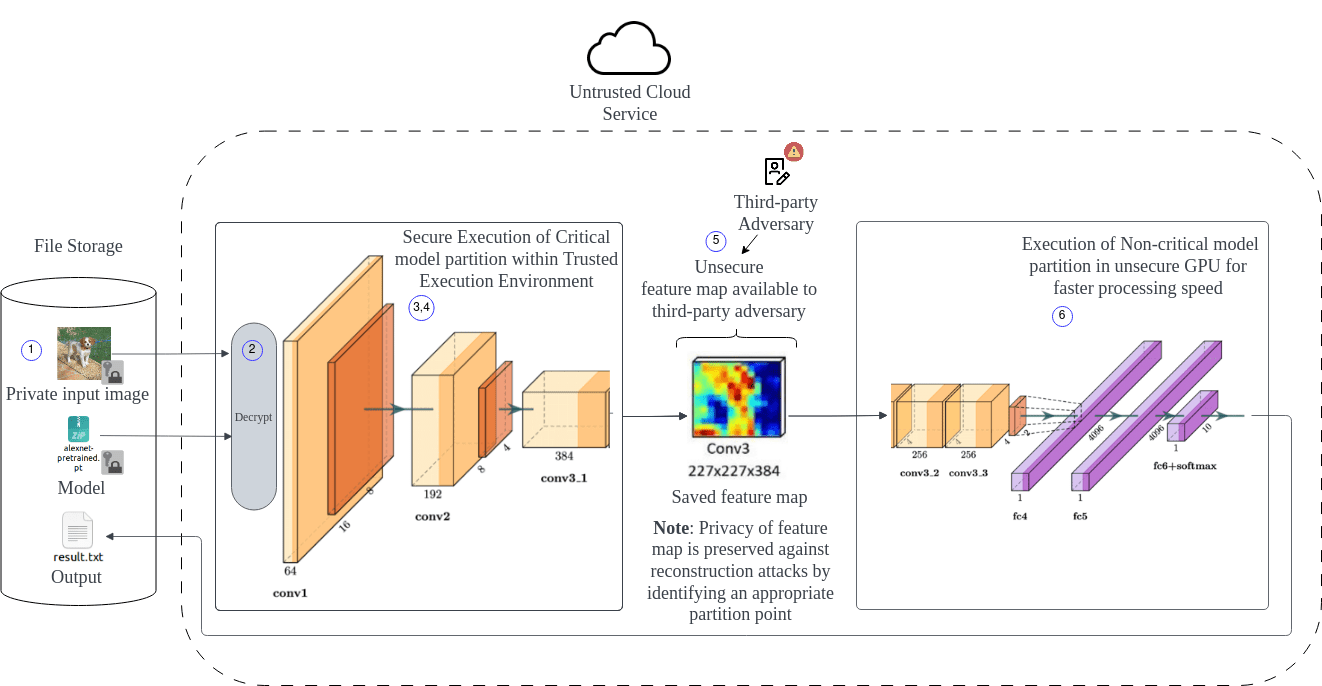}}
\caption{Secure system framework to perform private inference}
\label{framework}
\end{figure}

The framework of our solution is shown in Fig.~\ref{framework}. 
The steps for performing private inference using our solution are as follows:
\begin{enumerate}
    \item The user aims to utilize resources in an untrusted cloud for running their model and performing model inference on input data, all while ensuring the privacy of both the model and the data.
    \item The model and data are decrypted within the secure and private TEE which is hosted on the cloud. The information within the TEE cannot be exposed to the untrusted cloud environment. 
    \item The model is split into critical and non-critical partitions within the TEE, based on the architecture of the CNN model and the optimal partitioning point.
    \item The execution of the critical model partition on the input data is performed within the secure TEE.
    \item The output of the critical model partition which is saved and sent out into the untrusted cloud for further processing. 
    \item The intermediate feature maps and non-critical model partition can be loaded and executed in a cloud-based GPU for optimized runtime performance.
\end{enumerate}


\section{Privacy Preserving Inference Performance Evaluation}
\label{sec:eval}

In this work, we aim to explore the performance improvements and trade-off using model partitioning technique. This section consists of four parts: Neural network models, Datasets, Runtime performance evaluation and Privacy evaluation.
\subsection{Neural Network Models}
\label{sec:nnmodels}

\subsubsection{VGG-16} 
The architecture of VGG-16  \cite{simonyan2014very} is shown in Fig.~\ref{vgg_arch}. The model consists of 16 layers in total, which includes 13 convolutional layers and 3 fully connected (FC)
layers.





\subsubsection{ResNet-50} 
ResNet-50 is a CNN architecture \cite{he2016deep} composed of 50 layers, with its 16 residual blocks divided into four stages comprising 3, 4, 6 and 3 blocks as shown in Fig.~\ref{res50_arch}. Skip connections between blocks are used to mitigate the vanishing gradient problem.

\subsubsection{EfficientNetB0} 
The EfficientNetB0 architecture \cite{tan2019efficientnet}, as illustrated in Fig.~\ref{eff_arch}, is composed of 16 Mobile Inverted Bottleneck Convolution (MBConv) layers divided into seven stages composed of varying number of layers 
and one FC layer. 

\subsection{Datasets}
\label{sec:datasets}
\subsubsection{ImageNet} 
The ImageNet ILSVRC dataset\cite{imagenet} is a widely used collection of images belonging to 1000 classes for image classification tasks. 
\begin{figure}[htbp]
    \centering
        \begin{subfigure}[b]{0.09\textwidth}
            \includegraphics[width=\textwidth]{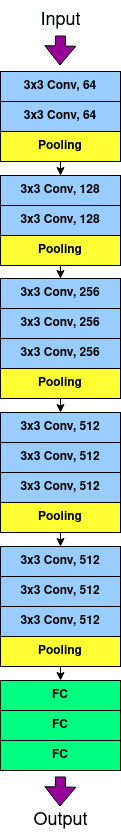}
            \caption{VGG-16}
            \label{vgg_arch}
        \end{subfigure}
        \begin{subfigure}[b]{0.16\textwidth}
            \includegraphics[width=\textwidth]{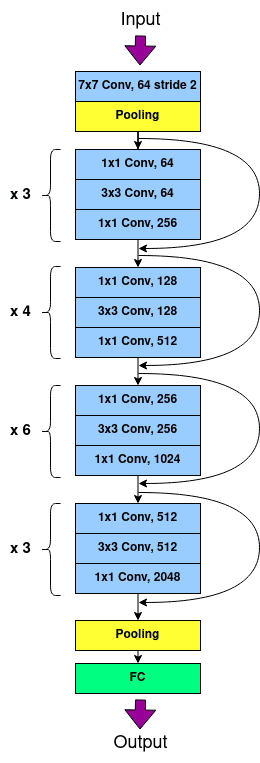}
            \caption{ResNet-50}
            \label{res50_arch}
        \end{subfigure}
        \begin{subfigure}[b]{0.15\textwidth}
            \includegraphics[width=\textwidth]{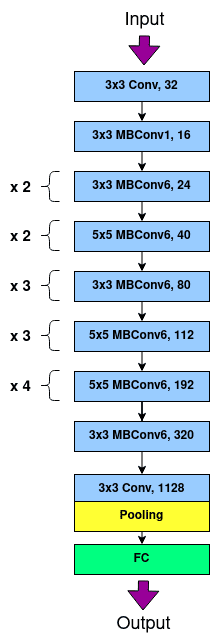}
            \caption{EfficientNetB0}
            \label{eff_arch}
        \end{subfigure}  
    \caption{Model architectures}
    \label{archs}
\end{figure}

\subsubsection{TON\_IoT Dataset converted to Images} 
In 2019, TON\_IoT Dataset \cite{alsaedi2020ton_iot} was created based on a testbed environment at the Cyber Range and IoT Labs at the University of New South Wales (UNSW) Canberra, Australia. The TON\_IoT dataset is converted to images by a processing method proposed in \cite{kodyvs2021intrusion} converting the time-based one-dimensional data in the original dataset\cite{alsaedi2020ton_iot} into tensors that are accepted by CNNs. 
The total number of classes are 8 attack types. For clarity, we will refer to the TON\_IoT Dataset converted to images as the "TON\_IoT image dataset".

\subsection{Runtime Performance Evaluation}
\label{sec:run}
In this study, we compared the runtimes of three models when offloading different layers to the GPU. 
The partitioning approach is used to help strike a balance between privacy preservation and computational efficiency, utilizing the functions of both Intel SGX and GPU acceleration. 
Gramine cannot directly access to the GPU from within the enclave. To address this limitation, we create a separate PyTorch process outside the enclave. This external process is responsible for offloading computations to the GPU. 
The kernel switch occurs only once during the offloading process and is measured as the time taken to save the feature map and load it onto the GPU. This operation typically takes between 0.02 to 0.1 seconds depending on the feature map size.
\subsubsection{Experimental Setup}
\label{sec:expset_run}
Our experiments measure the runtime for inference of the three aforementioned models offloading the intermediate feature maps at different partition points.
The inference within the TEE was done using Gramine library to run secure model inference within Intel SGX. 
The dataset used was a subset of 100 samples from the Imagenet dataset to measure the average inference runtimes of models partitioned at each layer. The runtimes measured are in seconds. The baseline used for each of the models is its runtime performance in Full-Enclave setting. The code to run inference, for critical layers in TEE and non-critical layers in GPU, was written in Python 3.8.
We conducted our evaluations on a desktop machine comprising an Intel® Core™ i7-9700K CPU with SGX capability, 8 threads and 64 GB of memory. We used an NVIDIA GeForce RTX 2070 Ti GPU 
as the accelerator. The operating system used is Ubuntu 22.04.2 LTS.

\subsubsection{Analysis}
\label{sec:runanalysis}

\paragraph{VGG-16}

\begin{figure}[htbp]
    \centering
    \begin{subfigure}[b]{0.59\textwidth}
        \begin{subfigure}[b]{0.27\textwidth}
            \includegraphics[width=\linewidth]{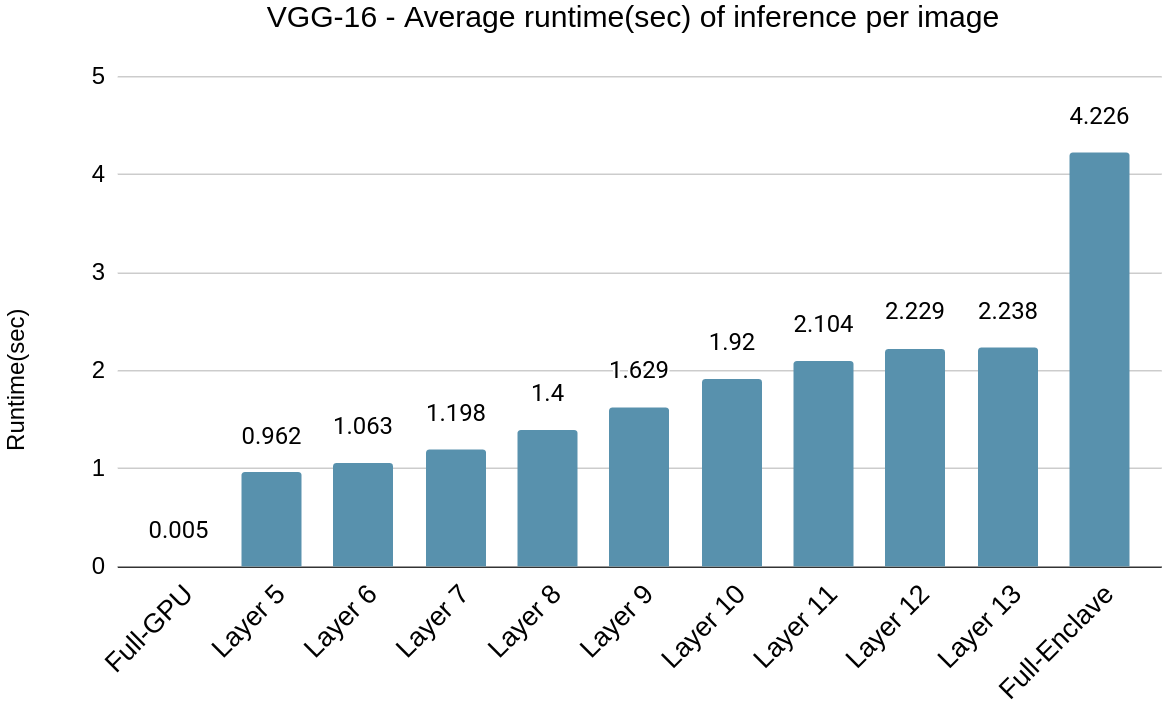}
            \caption{VGG-16}
            \label{vgg_run}
        \end{subfigure}
        \begin{subfigure}[b]{0.27\textwidth}
            \includegraphics[width=\linewidth]{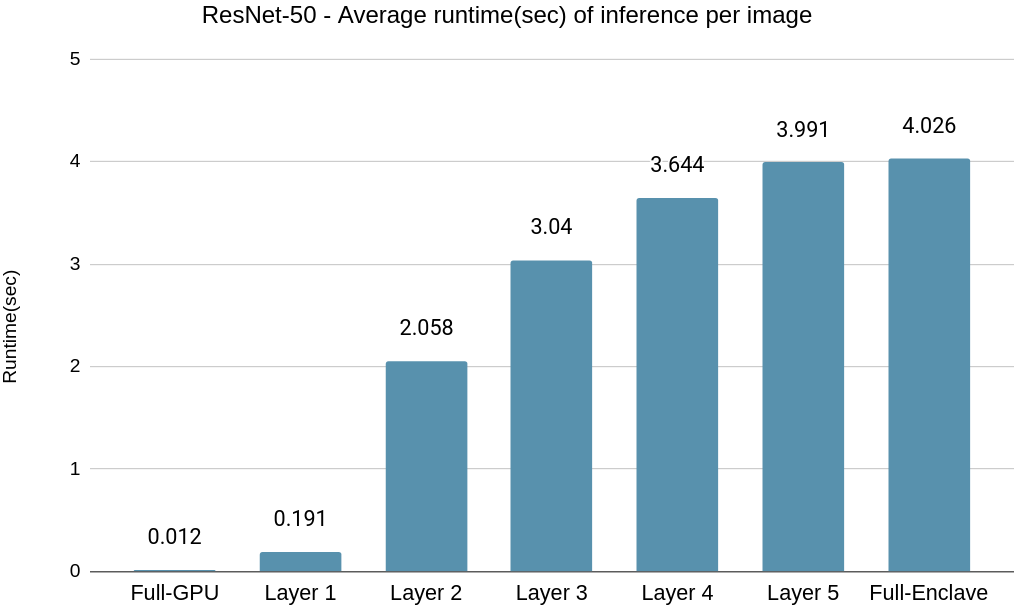}
            \caption{ResNet-50}
            \label{res50_run}
        \end{subfigure}
        \begin{subfigure}[b]{0.27\textwidth}
            \includegraphics[width=\linewidth]{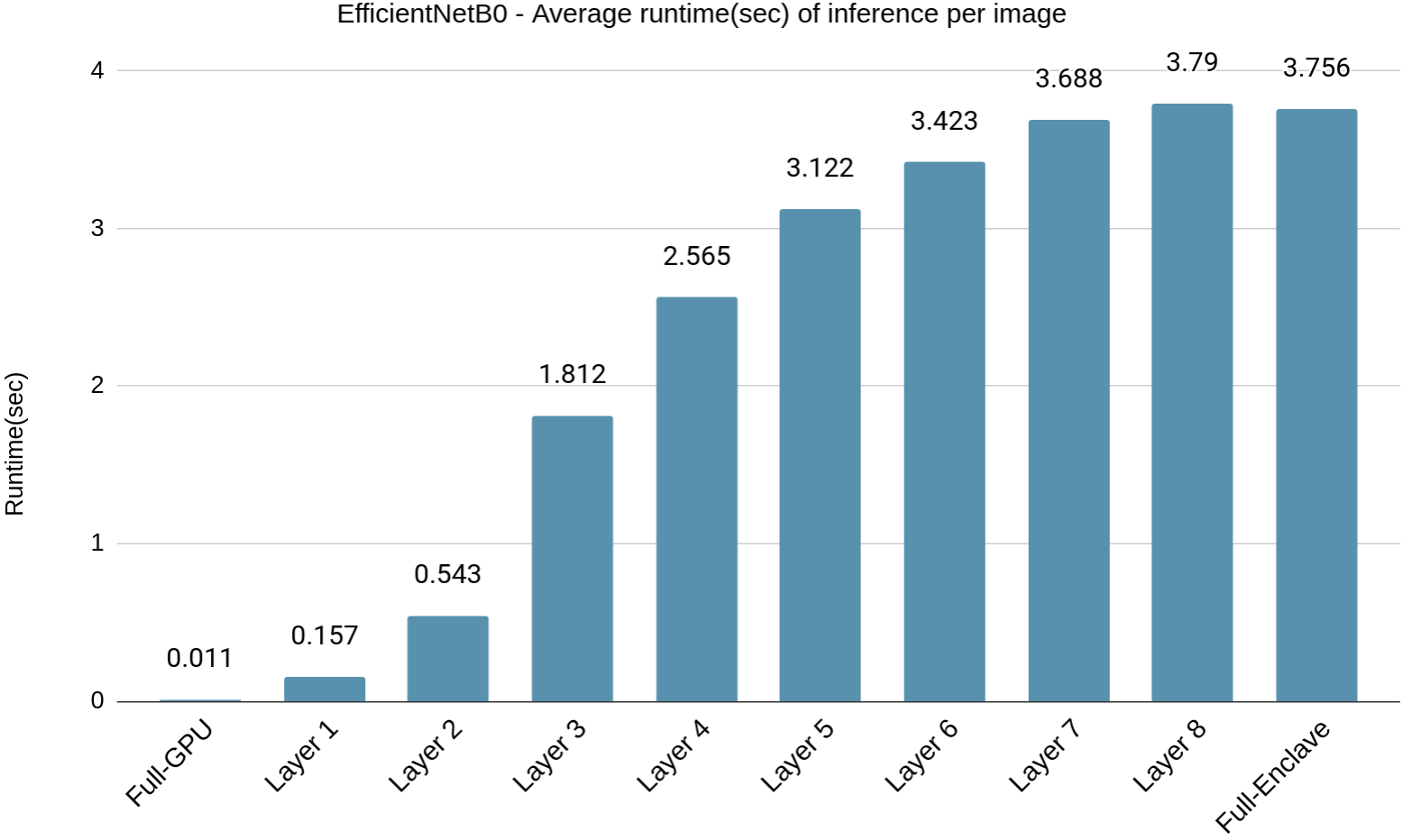}
            \caption{EfficientNetB0}
            \label{eff_run}
        \end{subfigure}
    \end{subfigure}


    \caption{Average inference runtime of DNN models for different layer partitions.}
    \label{modelruntimes}
\end{figure}

In Fig.~\ref{vgg_run}, we present the average inference runtime of the VGG-16 model for different layer partitions.
The x-axis represents the partitioning points. For instance, the x-axis label Layer 5 denotes the partitioning point at the 5th convolutional layer of the model from the input side which corresponds to the \texttt{3x3 Conv, 256} in VGG-16 architecture shown in Fig.~\ref{vgg_arch}. The partitioning is done after the convolutional layer in each instance. In Fig.~\ref{vgg_run}, it can be observed that there is a steady increase in runtime with the increase in the number of layers executed in enclave SGX.
\paragraph{ResNet-50} 
\label{sec:res50run}

In the case of ResNet-50, we observe the same trend of increased inference runtime when the model is partitioned at later layers, resulting in offloading execution of fewer layers to the GPU, as illustrated in Fig.~\ref{res50_run}.
The four stages in ResNet-50 contain various number of residual blocks, and each residual block contains 3 convolutional layers. For instance, the first stage contains 3 residual blocks, second stage contains 4 residual blocks. The partition is done at the end of each stage.
 Table \ref{res50table} shows the number of convolutional layers executed within the enclave SGX and GPU at different layer partitioning points.

\begin{table}[htbp]
\caption{ResNet-50 Layer partitions}
\begin{center}
\resizebox{\columnwidth}{!}{%
\begin{tabular}{|c|c|c|c|c|c|}
\hline
\textbf{\textit{Environment}}&\textbf{\textit{Layer 1}} & \textbf{\textit{Layer 2}} & \textbf{\textit{Layer 3}} & \textbf{\textit{Layer 4}} & \textbf{\textit{Layer 5}} \\
\hline
\textbf{\thead{Enclave SGX}}&1 conv layer & 10 conv layers&22 conv layers &40 conv layers &49 conv layers \\
\hline
\textbf{\thead{GPU}}& \thead{48 conv +\\ 1 FC layers} & \thead{39 conv +\\ 1 FC layers}& \thead{27 conv +\\ 1 FC layers} & \thead{9 conv +\\ 1 FC layers} &\thead{1 FC layer} \\
\hline
\end{tabular}
}
\label{res50table}
\end{center}
\end{table}

\paragraph{EfficientNetB0} 
The runtime performance of inference for each of these partition points shown in Fig.~\ref{eff_run} indicates a steady upward trend as the number of layers executed in GPU decreases. 
Table \ref{efftable} shows the number of MBConv blocks executed within the enclave SGX and GPU at different layer partitioning points.

\begin{table*}[htbp]
\small
\caption{EfficientNetB0 Layer partitions}
\resizebox{\textwidth}{!}{%
\begin{tabular}{|c|c|c|c|c|c|c|c|c|}
\hline
\textbf{\textit{Environment}}&\textbf{\textit{Layer 1}} & \textbf{\textit{Layer 2}} & \textbf{\textit{Layer 3}} & \textbf{\textit{Layer 4}} & \textbf{\textit{Layer 5}} & \textbf{\textit{Layer 6}} & \textbf{\textit{Layer 7}} & \textbf{\textit{Layer 8}} \\
\hline
\textbf{\textit{Enclave SGX}}&
\thead{1 conv layer} & 
\thead{1 conv +\\1 MBConv \\layers}& 
\thead{1 conv +\\3 MBConv \\layers} &
\thead{1 conv +\\5 MBConv \\layers} &
\thead{1 conv +\\8 MBConv \\layers} &
\thead{1 conv +\\11 MBConv \\layers} &
\thead{1 conv +\\15 MBConv \\layers}&
\thead{1 conv +\\16 MBConv + \\1 conv layers}  \\
\hline
\textbf{\textit{GPU}}& 
\thead{16 MBConv \\+ 1 conv +\\ 1 FC layers} & 
\thead{15 MBConv \\+ 1 conv +\\ 1 FC layers} & 
\thead{13 MBConv \\+ 1 conv +\\ 1 FC layers}& 
\thead{11 MBConv \\+ 1 conv +\\ 1 FC layers}&
\thead{8 MBConv \\+ 1 conv +\\ 1 FC layers}& 
\thead{5 MBConv \\+ 1 conv +\\ 1 FC layers}&
\thead{1 MBConv \\+ 1 conv +\\ 1 FC layers}&
1 FC Layer\\
\hline
\end{tabular}%
}
\label{efftable}
\end{table*}

\subsection{Privacy Evaluation: Measuring reconstruct-ability of intermediate feature maps}
\label{sec:privacy}
After analyzing the runtime performance, we evaluate the privacy of the offloaded feature maps at different partition points
determining the optimal layer partition for each model that balances speed-up and privacy. This is done by assessing the degree to which input images can be reconstructed from intermediate feature maps. 
Offloading more layers to the GPU speeds up inference but increases the risk of compromising input privacy. Striking the right balance is crucial to ensure privacy is not compromised.
In order to evaluate the privacy, we adopt the use of a c-GAN adversary model 
to reconstruct input images from the intermediate feature maps from different layer partitions. 
We evaluate all three models on ImageNet, and two models ResNet-50, EfficientB0 
(top performing models based on accuracy as per\cite{kodyvs2021intrusion})
on the TON\_IoT image dataset. These models were chosen to represent different architectural characteristics and complexity levels. 


\subsubsection{Experimental Setup}
To reconstruct the input images, we trained c-GAN models that takes the feature maps as input and generates corresponding image samples. 
By training the c-GAN on a dataset with 1000 images and their corresponding intermediate feature maps for 200 epochs, we aimed to achieve realistic and accurate image reconstructions. We used Python 3.8 for all our experiments to train c-GAN models and perform model partitioning using PyTorch deep learning framework, version 1.9.1. 
The training was conducted on the same desktop machine with specifications identical to those mentioned in Section~\ref{sec:expset_run}. 
We quantitatively evaluated the privacy of the models by computing the Structural Similarity Index Measure (SSIM) scores between the original images and their reconstructed versions from the feature maps of different layers. 
The SSIM metric provides a measure of structural similarity and perceptual quality between two images. It ranges from 0 to 1. A higher SSIM value indicates a high similarity, in terms of structural information, between the original image and the reconstructed image.
We selected the threshold SSIM score of 0.2 to achieve the goal of safeguarding input privacy considering the perceptual quality of the reconstructed image. Subsequently, we determined the optimal layer for partitioning to be the point at which the SSIM score falls below this threshold and consistently remains low thereafter. 

\subsubsection{Analysis of models on ImageNet dataset}
The summary results can be seen in Table \ref{analysis_sum}.

\begin{table}[htbp]
\Large
\caption{Summary analysis results of model speedups on ImageNet dataset}
\begin{center}
\resizebox{\columnwidth}{!}{%
\begin{tabular}{|c|c|c|c|c|c|}
\hline
\textbf{\thead{\large DNN Model}} & 
\textbf{\thead{\large Total \\ \large Partition Points}} & 
\textbf{\thead{\large Optimal Partitioning \\ \large Point}} & 
\textbf{\thead{\large Full-Enclave Inference \\ \large Runtime (Avg)}} & 
\textbf{\thead{\large Partitioned Inference \\ \large Runtime (Avg)}} & 
\textbf{\thead{\large Performance \\ \large Speedup (\%)}} \\
\hline
\textit{VGG-16} & 13  & Layer 8 & 4.2 sec & 1.4 sec & 66.6\% \\
\hline
\textit{ResNet-50} & 5  & Layer 4 & 4.02 sec & 3.6 sec & 10.4\% \\
\hline
\textit{EfficientNetB0} & 8  & Layer 4 & 3.7 sec & 2.5 sec & 32.4\% \\
\hline
\end{tabular}
}
\label{analysis_sum}
\end{center}
\end{table}

\begin{figure}[htbp]
    \centering
    \begin{subfigure}[b]{0.35\textwidth}
        \includegraphics[width=\textwidth]{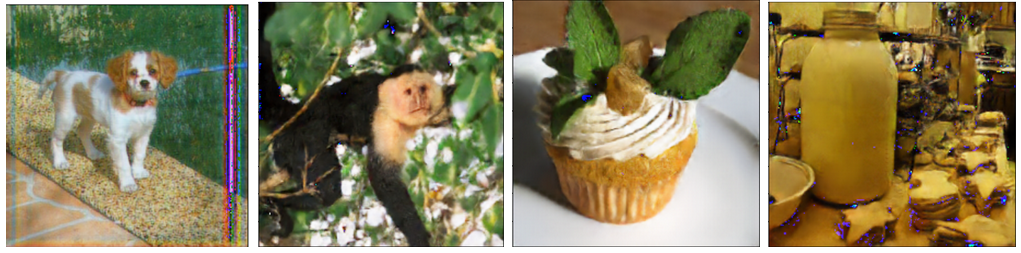}
        \caption{Reconstructed images from Layer 2 feature maps}
        \label{vgg2}
    \end{subfigure}
    \quad
    \begin{subfigure}[b]{0.35\textwidth}
        \includegraphics[width=\textwidth]{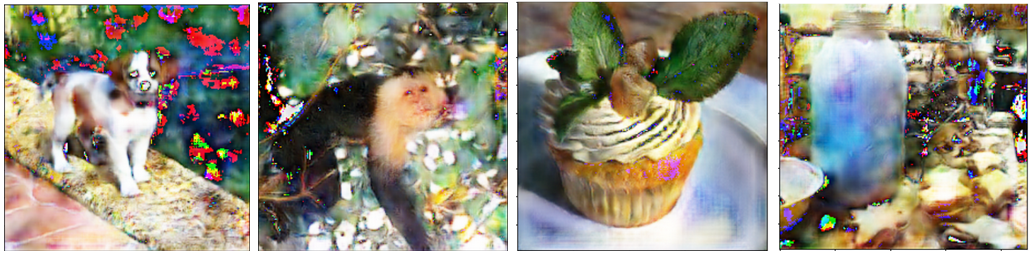}
        \caption{Reconstructed images from Layer 7 feature maps}
        \label{vgg7}
    \end{subfigure}
    \quad
        \quad
    \begin{subfigure}[b]{0.35\textwidth}
        \includegraphics[width=\textwidth]{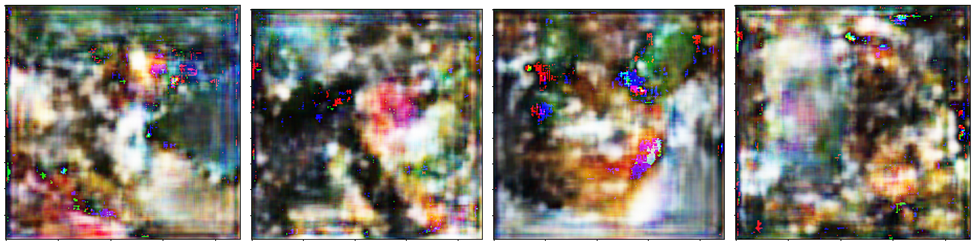}
        \caption{Reconstructed images from Layer 8 feature maps}
        \label{fig:vgg8}
    \end{subfigure}
    \caption{Reconstructed images from intermediate feature maps of different layer partitions in VGG-16}
    \label{vgg_rec}
\end{figure}

\begin{figure}[htbp]
    \centering

    \begin{subfigure}[b]{0.53\textwidth}
        \begin{subfigure}[b]{0.45\textwidth}
            \includegraphics[width=\linewidth]{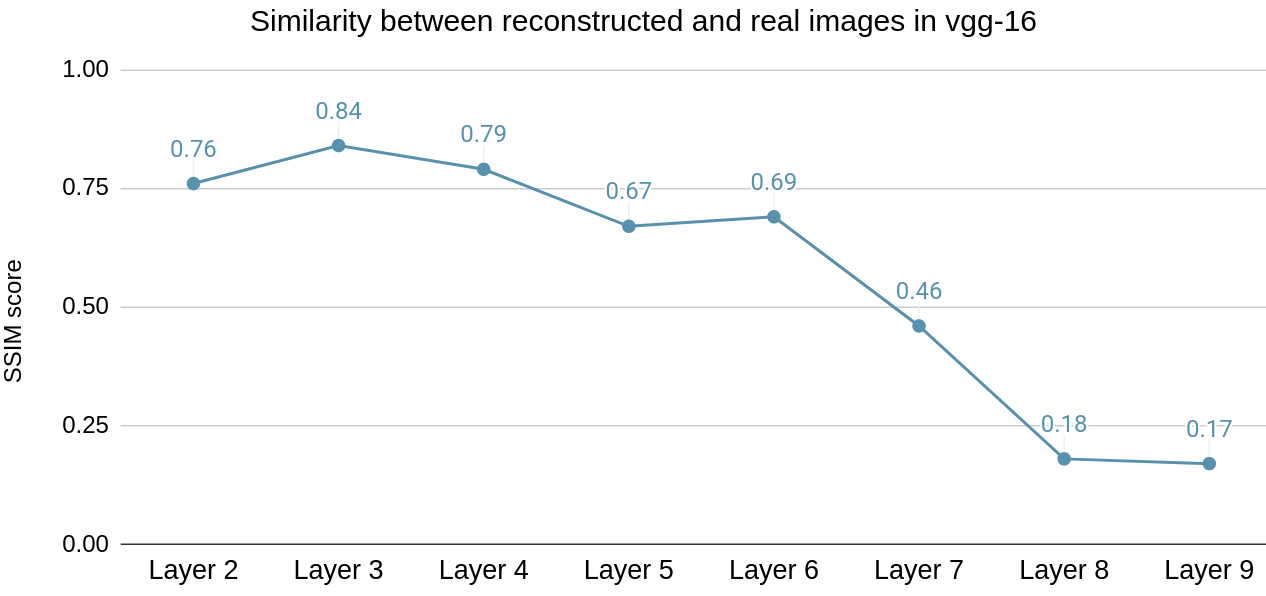}
            \caption{VGG-16}
            \label{vggss}
        \end{subfigure}
        \begin{subfigure}[b]{0.45\textwidth}
            \includegraphics[width=\linewidth]{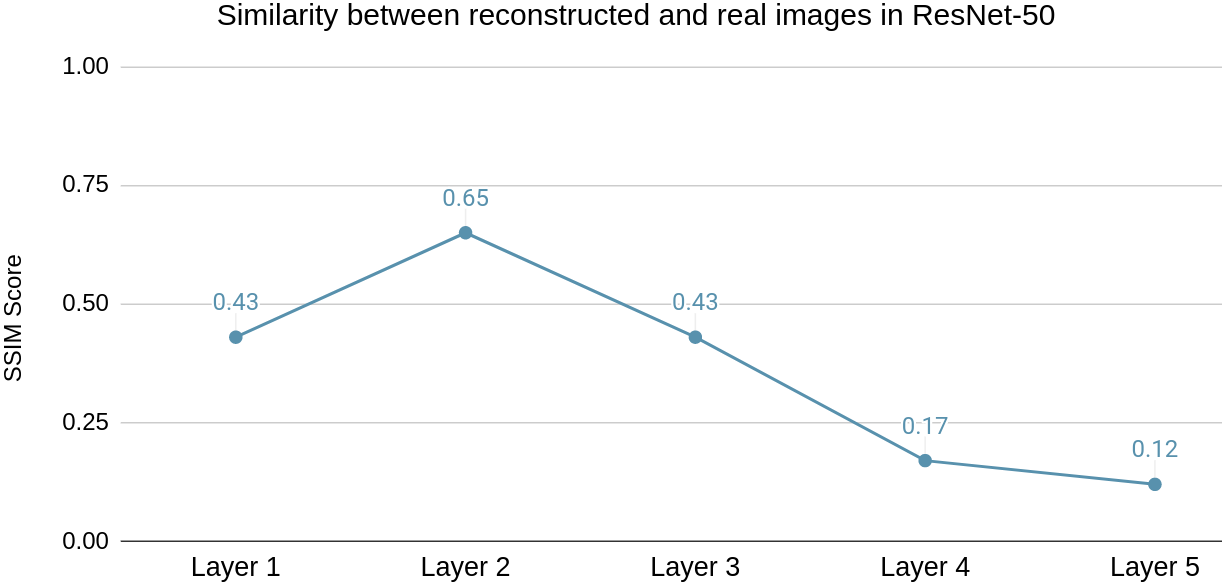}
            \caption{ResNet-50}
            \label{res50ss}
        \end{subfigure}
        \end{subfigure}


        \begin{subfigure}[b]{0.31\textwidth}

            \centerline{\includegraphics[width=\textwidth]{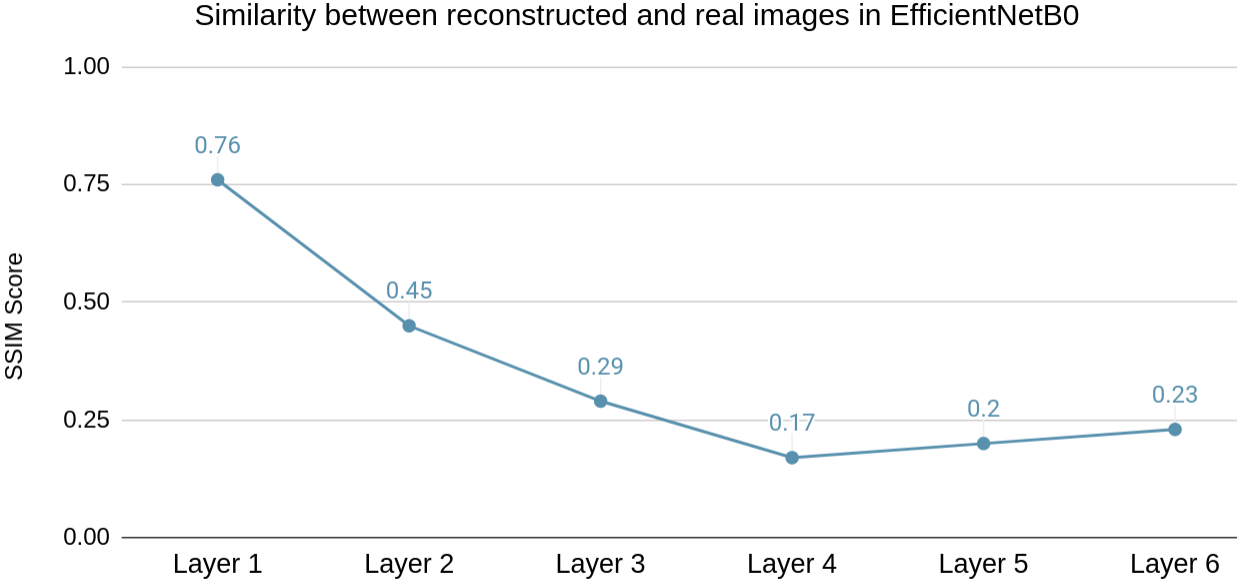}}
            \caption{EfficientNetB0}
            \label{effss}
        \end{subfigure}

    \caption{SSIM scores for reconstruction for different layer partitions of DNN models evaluated on ImageNet dataset.}
    \label{ssimsimg}
\end{figure}

\paragraph{VGG-16} 
Fig.~\ref{vggss} depicts the SSIM metric values for the similarity between the original and the reconstructed images obtained form c-GAN model Generator using the feature maps obtained from the respective layers.
It can be observed that there is a steady drop in the SSIM scores indicating that it becomes progressively more challenging to reconstruct the original images from the feature maps of the deeper layers. 
The SSIM score after Layer 7 remains below 0.2 and the reconstructed images can be seen in Fig.~\ref{vgg_rec} for further visual inspection. We can visually observe that the quality of reconstructed images from Layer 2 feature maps is superior to those from Layer 7 and Layer 8, indicating that it becomes increasingly challenging to reconstruct input images from deeper layers.

The SSIM score stays below 0.2 after Layer 8, indicating that Layer 8 is the optimal partitioning point. This decision is based on both the SSIM score metric and visual inspection of the reconstructed images. The average inference runtime is 1.4 seconds partitioned at Layer 8 compared to 4.2 seconds for Full-Enclave execution as shown in Fig.~\ref{vgg_run}. Hence, the speedup observed is 66.6\% compared to Full-Enclave inference.

\begin{figure}[htbp]
    \centering

    \begin{subfigure}[b]{0.4\textwidth}
        \centering
        \begin{subfigure}[b]{0.2\textwidth}
            \includegraphics[width=\textwidth]{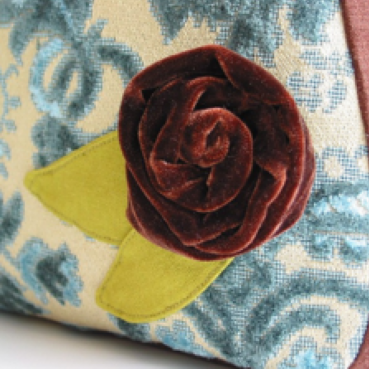}
        \end{subfigure}
        \begin{subfigure}[b]{0.2\textwidth}
            \includegraphics[width=\textwidth]{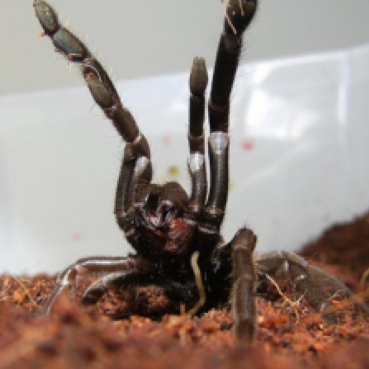}
        \end{subfigure}
        \begin{subfigure}[b]{0.2\textwidth}
            \includegraphics[width=\textwidth]{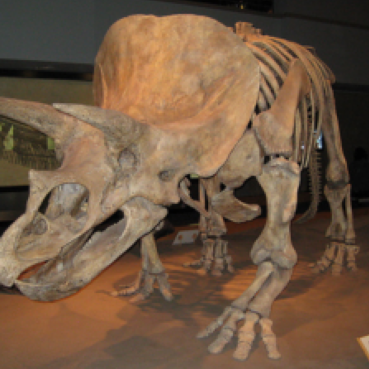}
        \end{subfigure}
        \begin{subfigure}[b]{0.2\textwidth}
            \includegraphics[width=\textwidth]{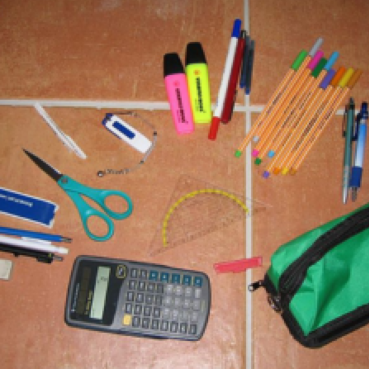}
        \end{subfigure}
        \caption{Original images}
        \label{res50_0}
    \end{subfigure}
    


    \begin{subfigure}[b]{0.4\textwidth}
        \centering
        \begin{subfigure}[b]{0.2\textwidth}
            \includegraphics[width=\textwidth]{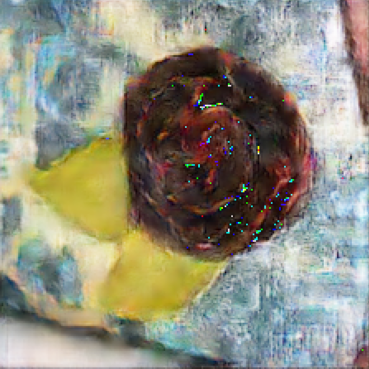}
        \end{subfigure}
        \begin{subfigure}[b]{0.2\textwidth}
            \includegraphics[width=\textwidth]{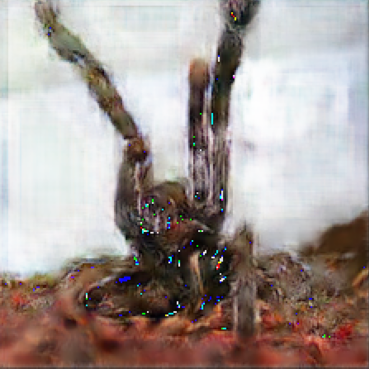}
        \end{subfigure}
        \begin{subfigure}[b]{0.2\textwidth}
            \includegraphics[width=\textwidth]{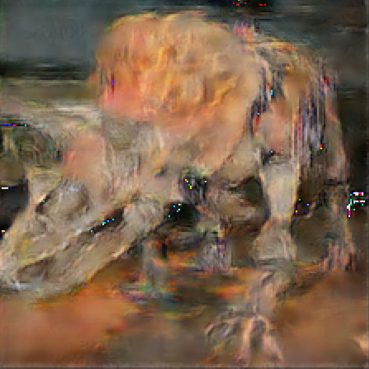}
        \end{subfigure}
        \begin{subfigure}[b]{0.2\textwidth}
            \includegraphics[width=\textwidth]{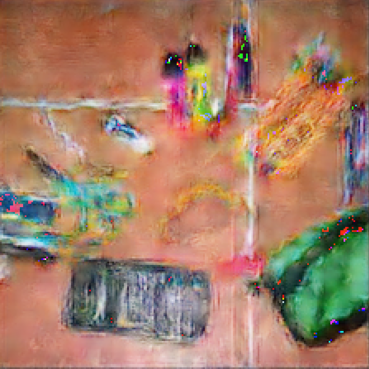}
        \end{subfigure}
        \caption{Reconstructed images from Layer 3 feature maps}
        \label{res50_3}
    \end{subfigure}

    \begin{subfigure}[b]{0.4\textwidth}
        \centering
        \begin{subfigure}[b]{0.2\textwidth}
            \includegraphics[width=\textwidth]{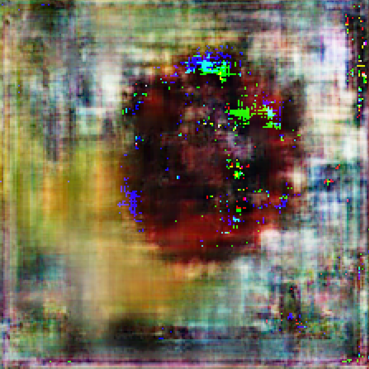}
        \end{subfigure}
        \begin{subfigure}[b]{0.2\textwidth}
            \includegraphics[width=\textwidth]{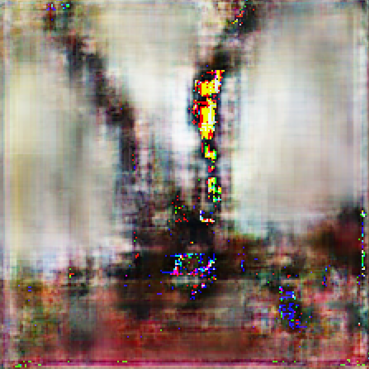}
        \end{subfigure}
        \begin{subfigure}[b]{0.2\textwidth}
            \includegraphics[width=\textwidth]{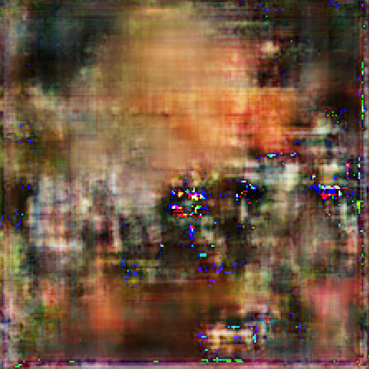}
        \end{subfigure}
        \begin{subfigure}[b]{0.2\textwidth}
            \includegraphics[width=\textwidth]{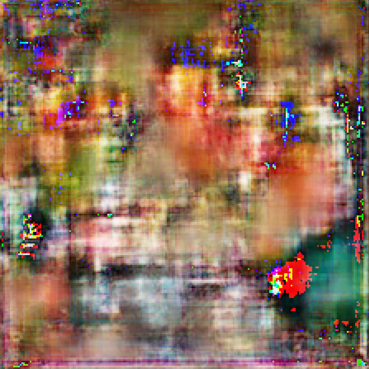}
        \end{subfigure}
        \caption{Reconstructed images from Layer 4 feature maps}
        \label{res50_4}
    \end{subfigure}

    
    \caption{Reconstructed images from intermediate feature maps of different layer partitions in ResNet-50.}
    \label{res50_recon}
\end{figure}

\paragraph{ResNet-50} 
ResNet-50 also follows the similar trend where it gets progressively harder to reconstruct images from feature maps of deeper layers. Here the SSIM score remains below 0.2 past Layer 3  as depicted in Fig.~\ref{res50ss}. 
Hence, Layer 4 is identified as the optimal layer to partition for ResNet-50 model inference based on SSIM score and visual inspection of Fig.~\ref{res50_recon} for the reconstructed images of ImageNet dataset. 
The average inference runtime is 3.6 seconds for model partitioned at Layer 4 compared to 4.02 seconds for Full-Enclave execution. Resulting in a speedup of 10.4\% compared to Full-Enclave inference. 

\paragraph{EfficientNetB0} 

\begin{figure}[htbp]
    \centering

    \begin{subfigure}[b]{0.4\textwidth}
        \centering
        \begin{subfigure}[b]{0.2\textwidth}
            \includegraphics[width=\textwidth]{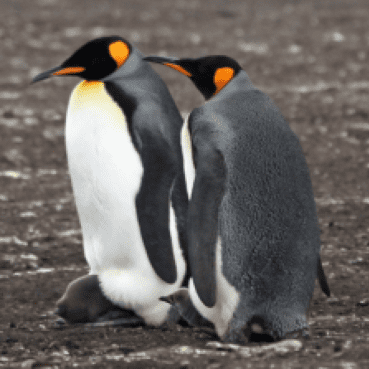}
        \end{subfigure}
        \begin{subfigure}[b]{0.2\textwidth}
            \includegraphics[width=\textwidth]{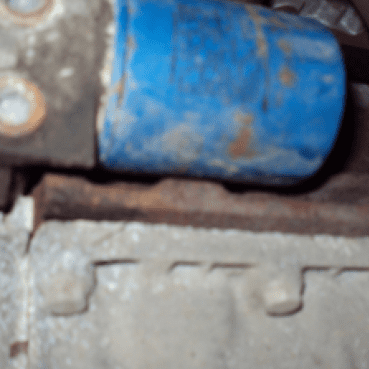}
        \end{subfigure}
        \begin{subfigure}[b]{0.2\textwidth}
            \includegraphics[width=\textwidth]{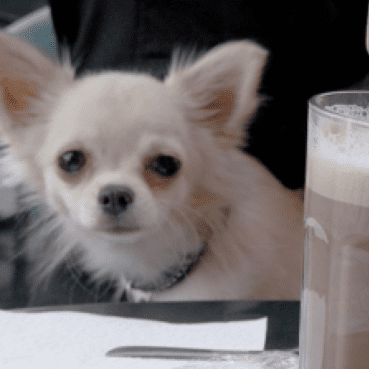}
        \end{subfigure}
        \begin{subfigure}[b]{0.2\textwidth}
            \includegraphics[width=\textwidth]{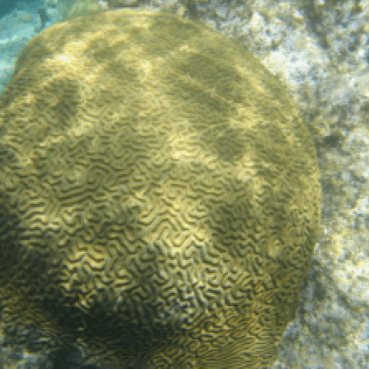}
        \end{subfigure}
        \caption{Original images}
        \label{eff_0}
    \end{subfigure}
    


    \begin{subfigure}[b]{0.4\textwidth}
        \centering
        \begin{subfigure}[b]{0.2\textwidth}
            \includegraphics[width=\textwidth]{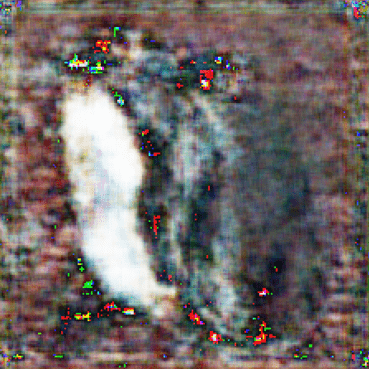}
        \end{subfigure}
        \begin{subfigure}[b]{0.2\textwidth}
            \includegraphics[width=\textwidth]{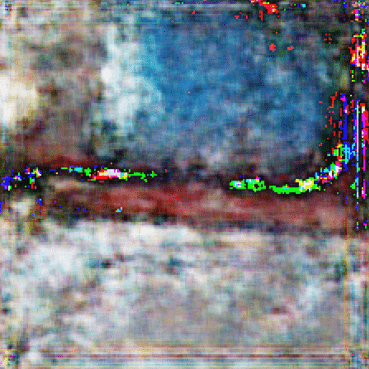}
        \end{subfigure}
        \begin{subfigure}[b]{0.2\textwidth}
            \includegraphics[width=\textwidth]{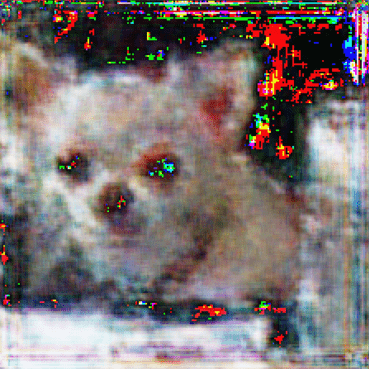}
        \end{subfigure}
        \begin{subfigure}[b]{0.2\textwidth}
            \includegraphics[width=\textwidth]{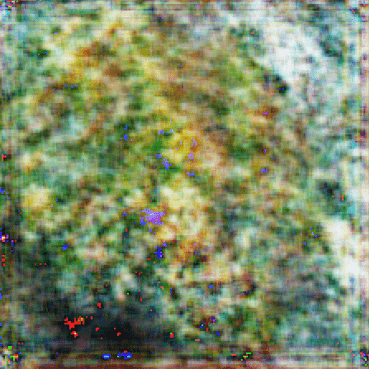}
        \end{subfigure}
        \caption{Reconstructed images from Layer 3 feature maps}
        \label{eff_3}
    \end{subfigure}

    \begin{subfigure}[b]{0.4\textwidth}
        \centering
        \begin{subfigure}[b]{0.2\textwidth}
            \includegraphics[width=\textwidth]{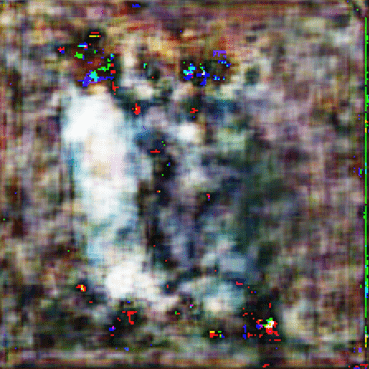}
        \end{subfigure}
        \begin{subfigure}[b]{0.2\textwidth}
            \includegraphics[width=\textwidth]{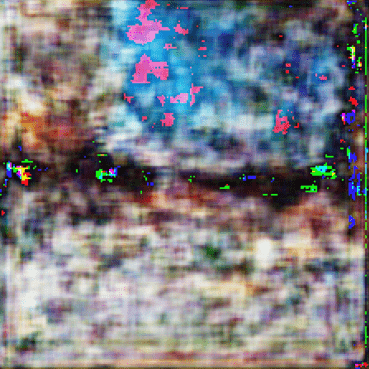}
        \end{subfigure}
        \begin{subfigure}[b]{0.2\textwidth}
            \includegraphics[width=\textwidth]{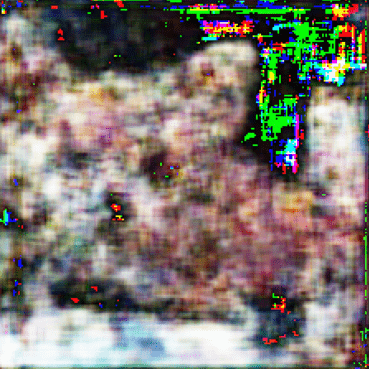}
        \end{subfigure}
        \begin{subfigure}[b]{0.2\textwidth}
            \includegraphics[width=\textwidth]{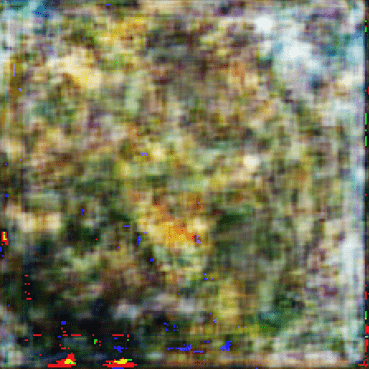}
        \end{subfigure}
        \caption{Reconstructed images from Layer 4 feature maps}
        \label{eff_4}
    \end{subfigure}

    \caption{Reconstructed images from intermediate feature maps of different layer partitions in EfficientB0.}
    \label{eff_recon}
\end{figure}

In the Fig.~\ref{effss}, it can be observed that there is a steady drop in the SSIM scores until Layer 4 and slight increase in the scores until Layer 6 of EfficientNetB0. However, it hovers around 0.2, which shows limited effectiveness of the reconstructed images from deeper layers.
The reconstructed images can be seen in Fig.~\ref{eff_recon} for further visual inspection.
The SSIM score from Layer 4 onwards remains around or below 0.2.  Hence the optimal layer to partition will be Layer 4. The average runtime for inference partitioning at Layer 4 of EfficientNetB0 is 2.5 seconds compared to 3.7 seconds for Full-Enclave execution. Hence, the speedup observed is 32.4\% compared to Full-Enclave inference.

\subsubsection{Analysis of models on TON\_IoT image dataset}



   


\paragraph{ResNet-50} 

\begin{figure}[htbp]
    \centering

    \begin{subfigure}[b]{0.53\textwidth}
        \begin{subfigure}[b]{0.45\textwidth}
            \includegraphics[width=\linewidth]{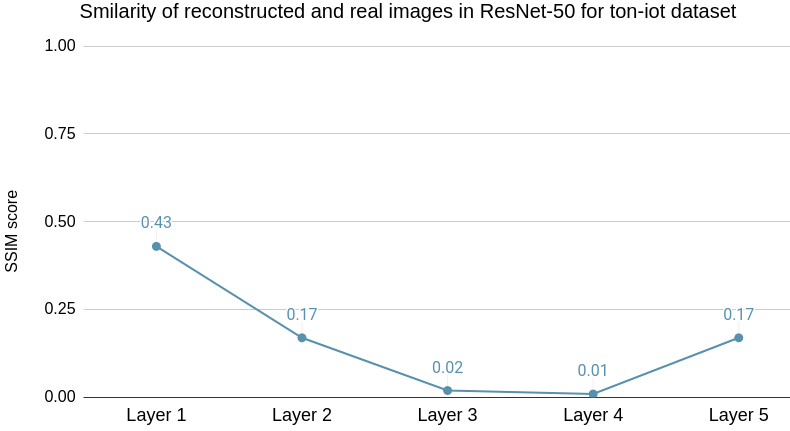}
            \caption{ResNet-50}
            \label{tonresss}
        \end{subfigure}
        \begin{subfigure}[b]{0.45\textwidth}
            \includegraphics[width=\linewidth]{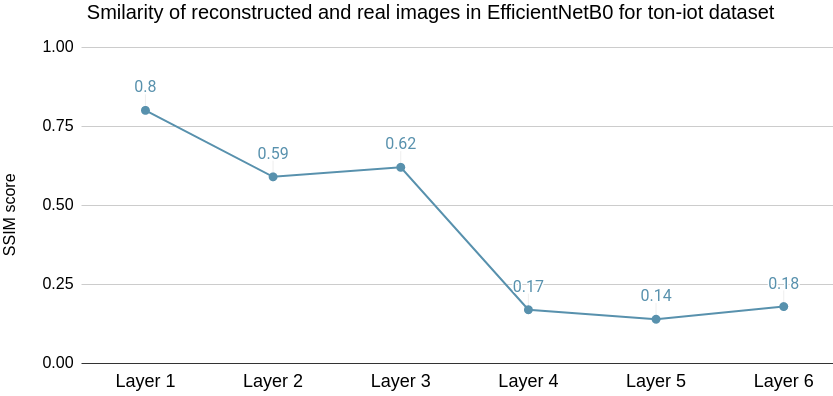}
            \caption{EfficientNetB0}
            \label{toneffss}
        \end{subfigure}
        \end{subfigure}
    \caption{SSIM scores for reconstruction for different layer partitions of ResNet-50 and EfficientNetB0 evaluated on TON\_IoT dataset.}
    \label{ssimston}
\end{figure}

The SSIM scores, in Fig.~\ref{tonresss} show a decreasing trend until Layer 4, followed by an increase at Layer 5 when evaluated on the TON\_IoT image dataset. 
It was observed that the reconstructions contain high noise from Layer 3 onwards.
Hence, the optimal layer to partition the model for use with TON\_IoT image dataset is Layer 3 whereas the optimal partition point for evaluation on ImageNet was identified to be Layer 4. 
The average inference runtime is 3.04 seconds for model partitioned at Layer 3 compared 4.02 seconds in Full-Enclave setting, resulting in a speedup of 24.3\% compared to Full-Enclave inference. 

\paragraph{EfficientNetB0} 

SSIM scores measured for the layers of EfficientNetB0 model for TON\_IoT image data can be seen in Fig.~\ref{toneffss}. 
The SSIM score for Layer 1, Layer 2 and Layer 3 was observed to be higher than that of SSIM scores for reconstruction based on the ImageNet dataset. However, the optimal layer to partition remains the same as Layer 4 as in the case with the ImageNet dataset where the average runtime is 2.5 seconds compared to 3.7 seconds for Full-Enclave execution. Hence, the speedup observed is 32.4\% compared to Full-Enclave inference.

\section{Conclusion}
\label{sec:conclusion}
Overall, our study showcases the potential of leveraging Intel SGX and GPU acceleration for privacy-preserving inference with deep learning models such as VGG-16, 
ResNet-50 and EfficientNetB0. We have conducted runtime performance analysis and assessed input privacy by measuring the input reconstruct-ability using trained c-GAN adversary models for each layer partition. The speedup rates differed among models, depending on their architecture and the choice of the optimal partition point. The majority of ML solutions and libraries are written in Python. To facilitate library efficient adoption, hence, we leverage Gramine as a Python package with seamless support for both Tensorflow and PyTorch. 
While Gramine allows running whole applications unmodified, it incurs high runtime costs for full execution of inference in enclave SGX. To address this, we use layer partitioning approach that improves performance while executing inference in TEE for critical layers and offloading rest of the computation for non-critical layers to a co-located GPU. 
As a future work, we plan to explore additional optimizations in performance for computationally intensive convolutional layers in the critical model partition
to further improve the inference runtime for models. 
\bibliographystyle{IEEEtran}
\bibliography{creferences} 

\end{document}